\documentclass{iau}

\usepackage{amsmath}
\usepackage{graphicx}
\usepackage{multirow}
\usepackage{hyperref}

\def\apj{{ApJ}}

\def\apjl{{ApJ Letters}} 
\def\apjs{{ApJ Supplement}} 
\def\apj{{ApJ}} 
\def\aap{{A\&A}}

\newcommand\prl{{Phys. Rev. Lett.}}%
\newcommand\ssr{{Space~Sci.~Rev.}}%

\newcommand{\solphys}{{\it Solar Phys.}}

\newcommand{\bmax}{$B_{\rm max}$}
\def\ndash{\,--\,}

\begin{document}

\lefttitle{B. K. Jha et al.}
\righttitle{BMRs Tilt Quenching}

\jnlPage{1}{7}
\jnlDoiYr{2021}
\doival{10.1017/xxxxx}

\aopheadtitle{Proceedings IAU Symposium}
\editors{A. Getling \&  L. Kitchatinov, eds.}

\title{Exploring the Quenching of Bipolar Magnetic Region Tilts using AutoTAB}

\author{Bibhuti Kumar Jha$^1$, Anu B. Sreedevi$^2$, Bidya Binay Karak$^2$ and\\ Dipankar Banerjee$^3$}
\affiliation{$^1$ Southwest Research Institute, Boulder 80302, USA\\ email:\email{maitraibibhu@gmail,com}}
\affiliation{$^2$Department of Physics, Indian Institute of Technology (Banaras Hindu University),\\ Varanasi 221005, India}
\affiliation{$^3$Aryabhatta Research Institute of Observational Sciences,\\ Nainital 263002, Uttarakhand, India}


\begin{abstract}
The tilt of the bipolar magnetic region (BMR) is crucial in the Babcock–Leighton process for the generation of the poloidal magnetic field in the Sun. We extend the work of \cite{Jha2020} and analyze the recently reported tracked BMR catalogue based on AutoTAB \citep{Sreedevi2023} from Michelson Doppler Imager (1996–2011) and Helioseismic and Magnetic Imager (2010–2018). Using the tracked information of BMRs based on AutoTAB, we confirm that the distribution of \bmax\ reported by \citet{Jha2020} is not because of the BMRs are picked multiple times at the different phases of their evolution instead it is also present if we consider each BMRs only once. Moreover, we 
find that the slope of Joy's law ($\langle\gamma_0\rangle$) initially increases slowly with the increase of \bmax. However, when \bmax$>$2.5\,kG, $\gamma_0$ decreases. The decrease of observed $\gamma_0$ with \bmax\ provides a hint to a nonlinear tilt quenching in the Babcock–Leighton process.

\end{abstract}


\maketitle

\section{Introduction}

The Sun and solar cycle variability have intrigued curious minds for over a century. Despite being enriched by continuous observations over the past century, solar cycle variability remains a primary puzzle in solar physics. However, it is now well established that the solar activity cycle is driven by the solar dynamo process, operating in the interior of the Sun, and it is responsible for the observed cyclic behavior of the Sun \citep{Charbonneau2014, Kar14a}. Solar differential rotation \citep{Jha2021, Jha2022} amplifies the magnetic field by twisting and stretching the existing polar field at the beginning of the solar cycle. These amplified and buoyant magnetic fields rise through the convection zone, where they experience the Coriolis force and emerge at the photosphere as tilted bipolar magnetic regions (BMRs). The residual flux from these BMRs migrates toward the poles and leads to the cancellation of existing opposite-polarity flux, initiating the onset of new flux buildup at the poles. These new opposite-polarity flux components act as the seed field for the following cycle and dictate its strength. The process of reversal and the buildup of new magnetic fields through the residual field carried to the poles by meridional flows is known as the Babcock-Leighton process \citep{Babcock1959, Babcock1961} .

One of the requirements for kinematic dynamo models, such as the Babcock-Leighton model, is the need for a nonlinear mechanism to suppress the exponential growth of the magnetic field \citep{Charbonneau2014}. 
Flux loss due to magnetic buoyancy through the formation of BMRs \citep{BKC22}, latitude quenching \citep{Kar20}, and the tilt quenching are the three potential nonlinearities identified in the solar dynamo \citep{Kar23}.
\citet{Lemerle2015} have first proposed that tilt quenching, which refers to the reduction of the tilt of Bipolar Magnetic Regions (BMRs) due to the presence of a strong magnetic field, can serve as the required nonlinearity in Babcock--Leighton dynamo models. Subsequently, \citet{Karak2017, Karak2018} incorporated tilt quenching in their dynamo models and achieved great success in reproducing the observed behavior of the solar cycle. More recently, \citet{Jha2020, Jha2023P} utilized magnetogram data from the Michelson Doppler Imager (MDI) and the Helioseismic and Magnetic Imager (HMI) to provide observational evidence of tilt quenching. Additionally, \citet{Jha2020} demonstrated that the distribution of the maximum magnetic field in BMRs exhibits a double peak in their distribution.

One limitation of the work by \citet{Jha2020, Jha2023P} is that they counted each BMR multiple times in their analysis, potentially impacting the inferences drawn from their results. To address this issue, \citet{Sreedevi2023, Sreedevi2023P} recently developed an algorithm based on feature association techniques to track BMRs in magnetograms more accurately. Here, we revisit the work done by \citet{Jha2020} and aim to determine whether the signature of tilt quenching reported by them can still be obtained from the tracked BMRs. Additionally, we investigate how the distribution of \bmax\ changes with the tracked information of BMRs

\section{Data and Method}
We use the line of sight (LOS) magnetograms from the MDI (1996\,--\,2011) and HMI (2011\,--\,2022) to identify the BMRs and track them through during the period when they are in the near side of the Sun. The detection algorithm consist of identification of strong magnetic field regions followed by a moderate flux balance condition to make sure the detected regions are BMRs.  This identification algorithm is similar to the one used in \citet{Stenflo2012a} and \citet{Jha2020}.  Recently \citet{Sreedevi2023} has taken another step and developed an automatic tracking algorithm for BMRs (AutoTAB) to track the BMRs identified by \citet{Jha2020}. The AutoTAB uses the features association technique, similar to one used by \citet{Jha2021} for sunspots tracking, to track the BMRs during its passage in the near side of the Sun. The AutoTAB provides a BMRs catalog with the properties of the BMRs such as  total flux, location, maximum field (\bmax), average field. Therefore, we use these properties of the BMRs to re-analyse the results obtained in \citet{Jha2020}.

\section{Results}
\subsection{Distribution of \bmax}
\begin{figure}[htbp!]
    \centering
    \includegraphics[width=\textwidth]{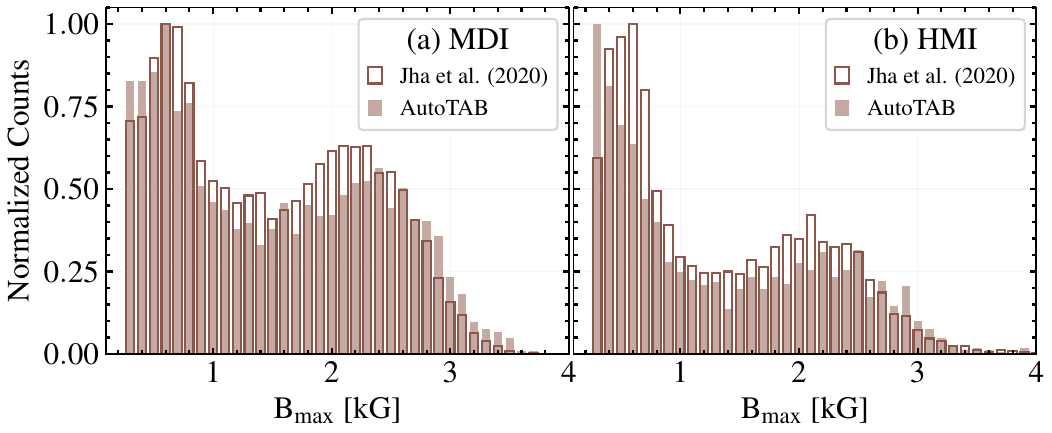}
    \caption{Normalized distribution of \bmax\ for BMRs tracked using AutoTAB represented by filled bars and \citet{Jha2020}, shown by unfilled bars.}
    \label{fig1}
\end{figure}

\citet{Jha2020} has reported that the \bmax\ show a bimodal distribution with one peaks close to 600\,G and another close to 2000\,G. These peaks get well seprated when they classified the BMRs based on their signature in white-light continuum images. They reported that the peak corresponding to 600\,G does not show 
any signature in white-light images i.e., they are not sunspots 
whereas other peak show a prominent signature for the same. This observed behaviour raised the question that, since in their work they have counted each BMRs multiple times it may be possible that they have picked the BMRs in their different phases and that gives rise to this observed double peak behavior. AutoTAB gives us an opportunity to track the BMRs and explore the distribution of \bmax\ in them. In Figure~\ref{fig1}, we show the distribution of \bmax, in which each BMRs are counted only once and the \bmax\ of BMRs is determined at the point of time where \bmax\ of BMRs peaks during their evolution. We noted that the distribution of \bmax\ in BMRs are indeed bimodal with peaks at 600\,G and 2\,kG. At this point we do not have the answer why the \bmax\ in BMRs show such distribution, it will be worth exploring in the future. In Figure~\ref{fig1} we also note that at many places the normalized fraction of BMRs exceed the fraction reported by \citet{Jha2020}. This is because  
the AutoTAB uses the data with higher cadence and for longer period. 

\subsection{Tilt Quenching}
\begin{figure}[htbp!]
    \centering
    \includegraphics[width=\textwidth]{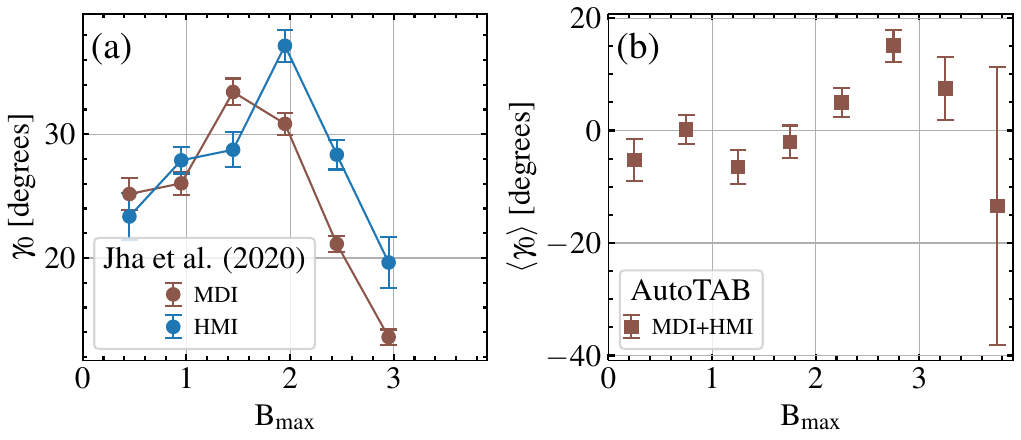}
    \caption{ (a) The amplitude of  Joy's law ($\gamma_0$) as a function of \bmax\ for MDI and HMI, taken from \citet{Jha2020}. (b) Average of amplitude of Joy's law ($\langle\gamma_0\rangle$) calculated over BMRs in 500\,G \bmax\ bin as a function of \bmax.}
    \label{fig2}
\end{figure}

Tilt quenching is the phenomena of reduction in tilt on top of Joy's law because of the stronger magnetic field in it. The idea of tilt quenching was given by \citet{DSilva1993}, where they have explained the origin of tilt in the BMRs. Late, \citet{Lemerle2015} and \cite{Karak2017, Karak2018} has used tilt quenching in their solar cycle model. 
\citet{Jha2020} 
gave the first direct evidence of tilt quenching in observation (Figure~\ref{fig2}(a)). In contrast with \citet{Jha2020}, where they have counted each BMR many times, here we used the AutoTAB to track the BMRs and use its tilt when the flux in the BMRs is maximum during their evolution. Tracking of BMRs has significantly reduced the number of data point and hence we can not look at the \bmax\ dependence of tilt in the way \cite{Jha2020} has looked at. Therefore we calculate the $\gamma_0$ of Joy's law which is $\gamma = \gamma_0\sin{\theta}$ for each BMR in each \bmax\ bin of size 500\,G. Here, $\gamma$ and $\theta$ is the tilt and latitude of BMRs, whereas $\gamma_0$ is called the amplitude of Joy's law. In Figure~\ref{fig2}(b), we show the average of $\gamma_0$ calculate in each \bmax\ bin along with the standard error as a function of \bmax. In Figure~\ref{fig2}(b) we note that the $\langle\gamma_0\rangle$ have quite different value compared to the Figure~\ref{fig2}(a), but still we can the similar trend in $\langle\gamma_0\rangle$ which show the downward trend after 2.5\,kG instead of 2.0\,kG. We emphasize here that this is the very preliminary result and detailed discussion will be reported in future publication.

\section{Conclusion}

In this article, we extend the work of \citet{Jha2020} by utilizing the newly developed BMR tracking algorithm AutoTAB \citep{Sreedevi2023}. AutoTAB was implemented on the LOS magnetogram data from MDI and HMI to track the BMRs during their passage on the near side of the Sun. The distribution of \bmax\ in BMRs shows a bimodal distribution similar to the one reported in \citet{Jha2020}, where they counted BMRs multiple times, unlike in this work. We also examine the variation of $\langle\gamma_0\rangle$ as a function of \bmax, showing a similar signature of tilt quenching as reported in \cite{Jha2020}. Since we calculate the average of $\gamma_0$ for each BMR in a 500,G \bmax\ bin, instead of fitting Joy's law, the values of $\langle\gamma_0\rangle$ differ from $\gamma_0$. These are preliminary results, and further details will be reported in an upcoming publication.

\bibliographystyle{iaulike}

\end{document}